# Controllable growth of 1-7 layers of graphene by chemical vapour deposition

Zhiqiang Tu,[a] Zhuchen Liu,[a] Yongfeng Li,[a,*] Fan Yang,[a] Liqiang Zhang,[a] Zhen Zhao,[a] Chunming Xu,[a] Shangfei Wu,[b] Hongwen Liu,[b] Haitao Yang,[b] and Pierre Richard[b]

[a]*State Key Laboratory of Heavy oil Processing, China University of Petroleum, Beijing Changping 102249, China*
[b]*Beijing National Laboratory for Condensed Matter Physics, Institute of Physics, Chinese Academy of Sciences, Beijing 100190，P. R. China*

**Abstract**

We report that graphene films with thickness ranging from 1 to 7 layers can be controllably synthesized on the surface of polycrystalline copper by a chemical vapour deposition method. The number of layers of graphene is controlled precisely by regulating the flow ratio of $CH_4$ and $H_2$, the reaction pressure, the temperature and the reaction time. The synthesized graphene films were characterized by scanning electron microscopy, transmission electron microscopy, selected area electron diffraction, X-ray diffraction and Raman spectroscopy. In addition, the graphene films transferred from copper to other substrates are found to have a good optical transmittance that makes them suitable for transparent conductive electrodes.

[*]Corresponding author. Tel: +86-10-89739028, E-mail: yfli@cup.edu.cn (Yongfeng Li)

# 1. Introduction

Graphene, a novel two-dimensional monolayer material consisting of a hexagonal crystal structure of $SP^2$-bonded carbon atoms, is gaining much attention due to its particular structure and great potential for a wide range of applications [1-4]. For instance, though graphene is as thin as one layer, it is also one of the strongest materials. Besides, enormous surface area of graphene with excellent conductivity and high carrier mobility can be grown. Literature survey shows that numerous methods have been developed to produce graphene, including the micromechanical cleavage of highly oriented pyrolytic graphite [1], epitaxial growth on a SiC substrate [5], reduction of graphite oxide [6], chemical vapour deposition (CVD) [7] and arc discharge [8]. Among all the developed synthesis methods, the CVD method has received significant attention for growing high-quality graphene films with large area. Up to now, this relatively simple and low-cost method has been used to produce graphene samples that can reach spectacular sizes [9], and can be easily transferred to other substrates [9, 10].

Interestingly, the various properties of materials made of more than one layer of graphene vary significantly with the number of layers, thus opening a new range of applications for carbon-based materials. Consequently, substantial effort is devoted to develop growth procedures for multi-layer graphene with controlled thickness that are compatible with manufacture technologies [11-13]. Although large area [14], excellent transparency [9] and large domain sizes [15, 16] of uniform graphene films have been obtained on Cu substrates [17] at low pressure, the control of the number of graphene layers for films with high uniformity is still not achieved due to the surface self-limited nature of carbon on Cu foil during the growth process [18, 19]. Moreover, in order to have a more comprehensive understanding of the optimization techniques for the synthesis of graphene by CVD, many attempts have been made for improving the uniformity and the quality of graphene, as well as for controlling precisely the number of layers using CVD graphene on Cu [16, 20-22].

Up to now, there are several reports about the controlled synthesis of few-layer graphene, which mainly focussed on the growth mechanism and the catalyst of graphene manufacture. Moreover, researchers prefer using Ni substrates owing to their larger carbon solubility, thus avoiding the problem of self-limitation. For example, Liu *et al.* reported a segregation approach to achieve precise layer control by using Cu-Ni alloy, which is a combination of the well-behaved Cu and Ni substrates for graphene growth [23]. Another method to synthesize few-layer graphene also based on Ni substrates, published by Gong *et al.*, is carried out by modulating simplified CVD process conditions and hydrogen exposure [24]. Herein, we employ a simple CVD method to investigate the relationship between the growth parameters and the number of layers of graphene, thus providing a promising approach to control the thickness of few-layer graphene films on Cu substrates. We also demonstrate the excellent electrical behaviour of our synthesized graphene films.

## 2. Experimental section

2.1 *Synthesis of 1-7 layers of graphene by a CVD method*

Graphene films with different number of layers were grown on 25-μm-thick copper foil (Alfa Aesar, item No.046365, cutting into 2×2 cm square plates) using the CVD method in a 4-inch chamber quartz reactor. In the case of the synthesis of the monolayer graphene film, a pretreatment by acetic acid and ethanol on the copper foil was first performed. Then the copper foil was sent to the centre of the furnace chamber. With argon flowing at a 300 sccm/min rate, the quartz tube was heated to 940 $^\circ$C, and the pressure was maintained at about $1.5\times10^3$ Pa. At this growth temperature, the flow ratio of $CH_4:H_2$ was maintained at 5:35 for 5 mins. After this step, $CH_4$ was cut off and the furnace was cooled down to room temperature under a 300 sccm/min flow of Ar and a 35 sccm/min flow of $H_2$. The heating rate was 12 $^\circ$C/min, while the cooling rate was 10 $^\circ$C/min.

Similarly, the other layered graphene films (2-7 layers) on Cu were fabricated by changing the synthesis recipe, as shown in Table 1. For instance, in order to obtain the bilayer graphene film, the flow ratio of $CH_4:H_2$ was changed to 10:30, while the other experiment conditions remained the same as for the growth of monolayer graphene. In contrast to the synthesis process for monolayer and bilayer graphene, the preparation of 3-7 layers graphene films was carried out under atmospheric pressure, and the flow ratio of $CH_4:H_2$ was the same as for the bilayer synthesis. The reaction temperature for trilayer graphene growth was 960 °C, and the reaction time was 5 mins. When synthesizing the four-layer and five-layer graphene films, the optimal temperature was found to be 920 °C, while the reaction times were 10 mins and 20 mins, respectively. The six and seven layers graphene films were synthesized at 940 °C, and the reaction times were 10 mins and 20 mins, respectively, which could be controlled similarly to the case of the four- and five-layer graphene films.

2.2. *Graphene transfer onto Si substrate*

Graphene films with different number of layers were obtained by etching Cu foil in an aqueous solution of ferric chloride. Typically, the graphene films grown on the copper foil were coated with poly-methyl methacrylate (PMMA) before etching. After the etching process, the films were transferred to deionized water. The PMMA/graphene stack was then lifted by the silicon. In order to smooth the wrinkles caused during the transfer process, ethanol was sprayed to the surface of the stack, which could make a full contact between PMMA/graphene and silicon. Water molecules were vaporized by drying the films in an oven at 50 °C for about 3 hours. Afterwards, acetone was used to dissolve the PMMA and clean graphene on silicon was finally obtained.

## 3. Results and discussion

The growth of graphene films with different numbers of layers using the CVD method on copper foil has been confirmed by scanning electron microscopy (SEM), transmission electron microscopy (TEM) and selected area electron diffraction (SAED). SEM, used to obtain images of graphene on Cu substrate, permits visualization of the location and shape of the graphene domains, as shown in Fig. 1a. Fig. 1b depicts a high-magnification SEM image of a synthesized graphene film transferred to a Si surface. According to the SEM images, the graphene films are continuous, although there are a number of wrinkles denoted by white areas. As shown in Figs. 2-3, TEM examination provides an accurate way to measure the number of layers in the few-layer graphene films produced according to the growth conditions summarized in Table 1. The low- and high-magnification TEM images of a monolayer graphene film transferred onto the TEM grid are shown in Fig. 2, along with its corresponding SAED pattern (inset). While a single layer can be identified from the TEM image displayed in Fig. 2b, the SAED pattern shown in inset reveals the typical hexagonal crystalline nature of graphene. Figs. 3a-f show the TEM images of few-layer graphene films (2 to 7 layers) for which the number of layers can be clearly determined. The SAED patterns associated to bilayer graphene, trilayer graphene and six-layer graphene films, which have been recorded at the centre of graphene domains to avoid regions containing back-folded edges, are displayed in Figs. 3g-I, respectively. In addition the SAED patterns of four-layer, five-layer, and seven-layer graphene films are very similar to case of six-layer graphene. We also found that due to the contribution from 2 layers, the SAED pattern of bilayer graphene contains 12 spots in each ring rather than just 6 as in monolayer graphene.

Raman spectroscopy is a nondestructive technique providing a quick and easy structural characterization. Fig. 4a shows the Raman spectrum of a monolayer graphene film on Si. The two most prominent features are the G band at 1584.9 cm$^{-1}$ and the 2D band at 2686.2 cm$^{-1}$. The G band corresponds to a $E_{2g}$ vibration mode at the zone centre common to sp$^2$ carbon materials. In addition to the G band and 2D band peaks, we observe a low intensity peak at ~1346.3 cm$^{-1}$ corresponding to the D band, which is associated with structural defects. While the G band is a symmetry-allowed Raman transition of graphite, the D band is ascribed to a double resonance (DR) Raman scattering process. The $I_D/I_G$ intensity ratio (0.161) is small in our sample, which is consistent with large size graphene with good quality [25]. Some weak D band intensity is also observed away from the graphene edges, which demonstrates the existence of subdomain boundaries in areas with a constant number of graphene layers. The spectra from the thinnest sections of the CVD graphene films show a sharp line width (~35 cm$^{-1}$) characteristic of monolayer graphene. The $I_{2D}/I_G$ intensity ratio, known to strongly correlate with the number of graphene layers [8], is large (2.4) in our samples, which is consistent with our TEM observation of a single graphene layer.

As illustrated in Fig. 4b, the Raman spectra of the few-layer graphene films transferred on Si substrates evolve as a function of the number of graphene layers. Obviously, as the number of layers increases from monolayer to multilayer, the 2D peak broadens and the G peak becomes more pronounced. In addition, there is a slight blue shift in the 2D band (from 2699.3 cm$^{-1}$ to 2686.2 cm$^{-1}$) and a gradual decrease in the $I_{2D}/I_G$ intensity ratio, which is consistent with previous studies [8, 15]. Based on previous studies [15] of CVD-grown graphene (transferred from Cu) and our current data, we conclude that $I_{2D}/I_G$ (2.4) >2 for monolayer graphene, 1< $I_{2D}/I_G$ (1.1) <2 for bilayer graphene, and $I_{2D}/I_G$ (0.4) <1 is a signature of trilayer graphene. As indicated in Fig. 4c though, the $I_{2D}/I_G$ intensity ratio varies little for thicker samples.

In addition, XRD measurements were performed for the synthesized graphene transferred on a glass sheet. In Fig. 5, the XRD spectrum of a trilayer graphene film presents an obvious peak. The XRD pattern of the as-prepared graphene shows a broad peak at 2θ = 25.25°, corresponding to inter-planar spacing of 9.48 Å, which is the reflection (002) of graphene [26]. The fairly broad reflection (002) suggests that the samples are very poorly ordered along the stacking direction. This result indicates that the graphene sheets stack into multilayers.

High transparency is also necessary for the use of CVD graphene as a substitute for transparent metal oxide electrodes in organic photovoltaic cells. The optical transmittance of the transferred graphene films in the visible and near infrared range was measured by using a UV756CRT spectrophotometer in the 400-1100 nm wavelength range. Because it is a π-conjugated system, light transmission in graphene is higher in the ultraviolet-visible range. As indicated in Fig. 6a, the light transmission of the films increases rapidly from 400 nm to about 550 nm, value above which the transmission almost saturates. As expected, the transparency of the few-layer graphene films decreases with the thickness of the films. We display in Fig. 6b the transmittance of the few-layer graphene films at wavenumber λ = 550 nm. While it decreases only from 93% to 92% for the first 3 layers of graphene, the optical transmittance shows a sharp decrease when the thickness of the films reaches four layers, and for seven layers the optical transmittance has dropped to 75%.

To the best of our knowledge, the CVD growth of graphene on Cu substrate is basically attributed to the low solubility of carbon (< 0.001 at.%) in Cu and to the surface diffusion of carbon atoms on Cu.

As a consequence, it is quite common to see the preferential growth of monolayer graphene, which can be explained by a self-limiting mechanism [14, 19]. However, our results confirm that graphene with 1-7 layers can be controllably synthesized on Cu substrates by the CVD method. Although the detail mechanism of formation remains unclear, the number of layers of graphene should be determined by the diffusion rate of carbon atoms, which is mainly affected by the concentration of the available carbon atoms on the Cu surface. Hence, in order to precisely control the number of graphene layers, more attention should be paid to control the concentration of the carbon atoms on the Cu surface. In our case, low pressure is favourable to the formation of a few layers (1-3) of graphene whereas atmosphere pressure is suitable for the formation of multilayer (3-7) graphene, suggesting that the carbon concentration and the diffusion rate play important roles in determining the number of layers of graphene. In addition, since the concentration of the carbon atoms relies on the partial pressure of the hydrocarbon gas at a given temperature, controlling the partial pressure of both the carbon sources and the $H_2$ gas is also critical for controlling the thickness and the quality of few-layer graphene films.

## 4. Conclusion

In conclusion, we have successfully developed a controllable technique to precisely synthesize 1 to 7 layers of highly uniform graphene films by regulating the flow ratio of $CH_4$ and $H_2$, the reaction the pressure, the temperature and the reaction time. The synthesis of few-layer graphene has been confirmed by SEM, TEM, SAED, XRD and Raman spectroscopy characterizations. The synthesis of graphene on Cu foil shows many advantages over the well-studied growth of graphene on Ni, including its easy controllability, low cost, easy transfer and good potential for large-scale production. Additionally, our new method opens a feasible route for the synthesis of wafer-scale graphene with a controllable number of layers, and it provides a new understanding of the graphene formation on Cu catalyst. Further work is in progress to apply the synthesized graphene to photoelectric devices.

## Acknowledgements

This work was financially supported by the National Natural Science Foundation of China (Nos. 51071175, 21106184 and 11274362), Science Foundation of China University of Petroleum, Beijing (Nos. 2462013YJRC40, YJRC-2011-18) and Thousand Talents Program.

Table 1. Summary of the growth conditions of few-layer graphene films

| Layers | Temperature (℃) | Pressure (Pa) | Time (Min) | $CH_4:H_2$ (sccm) |
|---|---|---|---|---|
| Monolayer | 940 | $1.5 \times 10^3$ | 5 | 5:35 |
| Bilayer | 940 | $1.5 \times 10^3$ | 5 | 10:30 |
| Three layers | 960 | $1.01 \times 10^5$ | 5 | 10:30 |
| Four layers | 920 | $1.01 \times 10^5$ | 10 | 10:30 |
| Five layers | 920 | $1.01 \times 10^5$ | 20 | 10:30 |
| Six layers | 940 | $1.01 \times 10^5$ | 10 | 10:30 |
| Seven layers | 940 | $1.01 \times 10^5$ | 20 | 10:30 |

**Figures captions**

Fig. 1. (a) SEM image of a graphene sheet on a copper substrate. (b) SEM image of a graphene sheet transferred on a Si substrate, showing several wrinkles.

Fig. 2. (a) Low-magnification TEM image of monolayer graphene. (b) High-magnification TEM image of monolayer graphene, where a SAED pattern recorded from the centre of the domains is shown in the inset.

Fig. 3. High-resolution TEM images of representative edges of graphene with different numbers of layers, (a) bilayer, (b) trilayer, (c) four layers, (d) five layers, (e) six layers and (f) seven layers (scale bar: 5nm). The typical SAED images of bilayer, trilayer and multilayer graphene taken from the centre of the domains are shown in (g) - (i), respectively.

Fig. 4. (a) Raman spectrum of monolayer graphene with the G peak observed at 1584.9 cm$^{-1}$, the D-peak observed at 1346.3 cm$^{-1}$ and the 2D peak observed at 2700 cm$^{-1}$. (b) Raman spectra of graphene with 1-7 layers, indicating that the intensities of the G peak and 2D peak are dependent on the number of layers. (c) The $I_{2D}/I_G$ intensity ratio as a function of the number of graphene layers, indicating that the ratio decreases slightly when the number of layers reaches three.

Fig. 5. XRD curve of trilayer graphene, with a broad peak at $2\theta = 25.25°$ suggesting that the graphene film is very poorly ordered along the stacking direction, and that the graphene sheets stack into multilayers.

Fig. 6. Optical properties of CVD graphene, (a) UV-visible transmittance spectra of few-layer graphene films on quartz substrates, indicating that the transmittance becomes smaller as the number of layers increases, (b) The transmittance at 550 nm versus the number of layers further enhances the comparison in the optical transparency of the films.

**Figure 1**

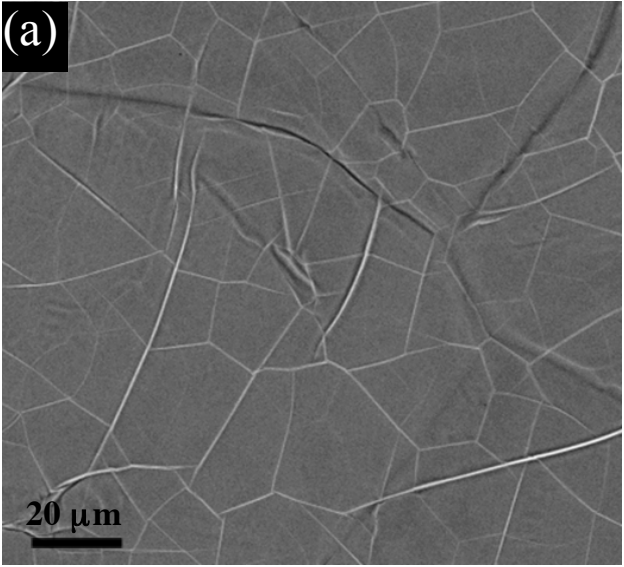

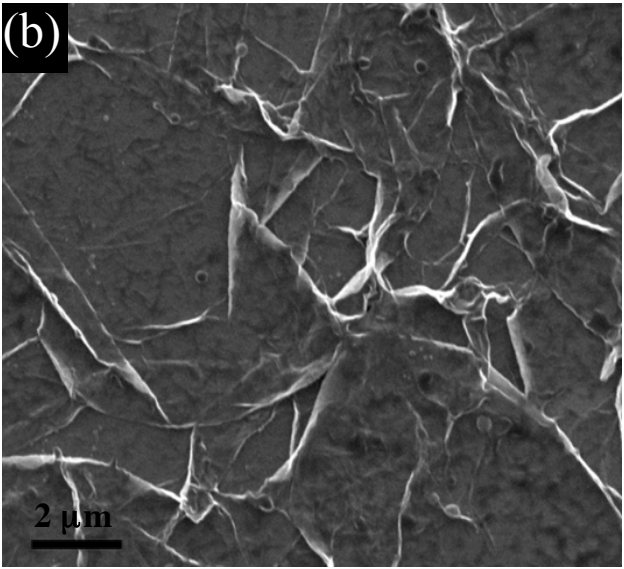

**Figure 2**

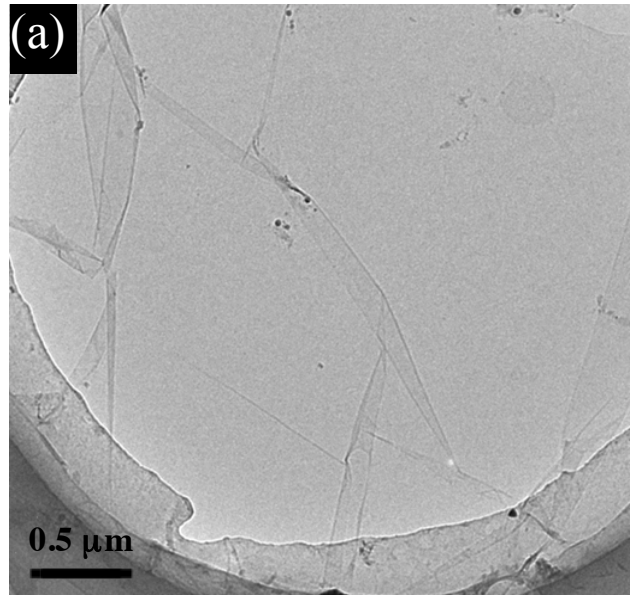

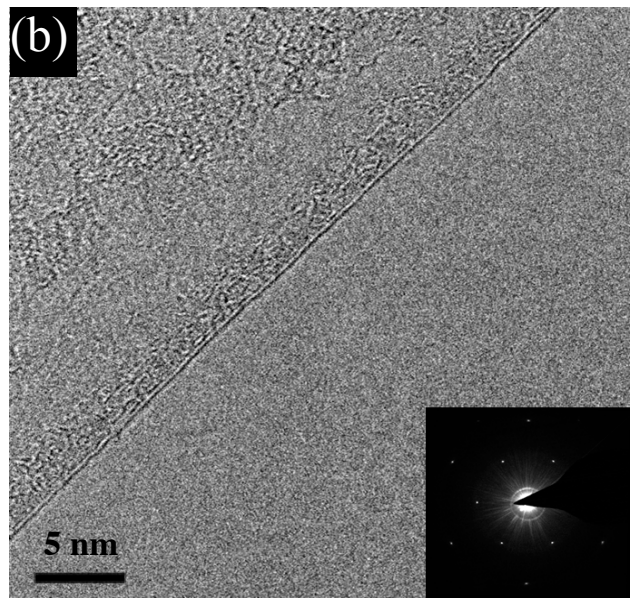

**Figure 3**

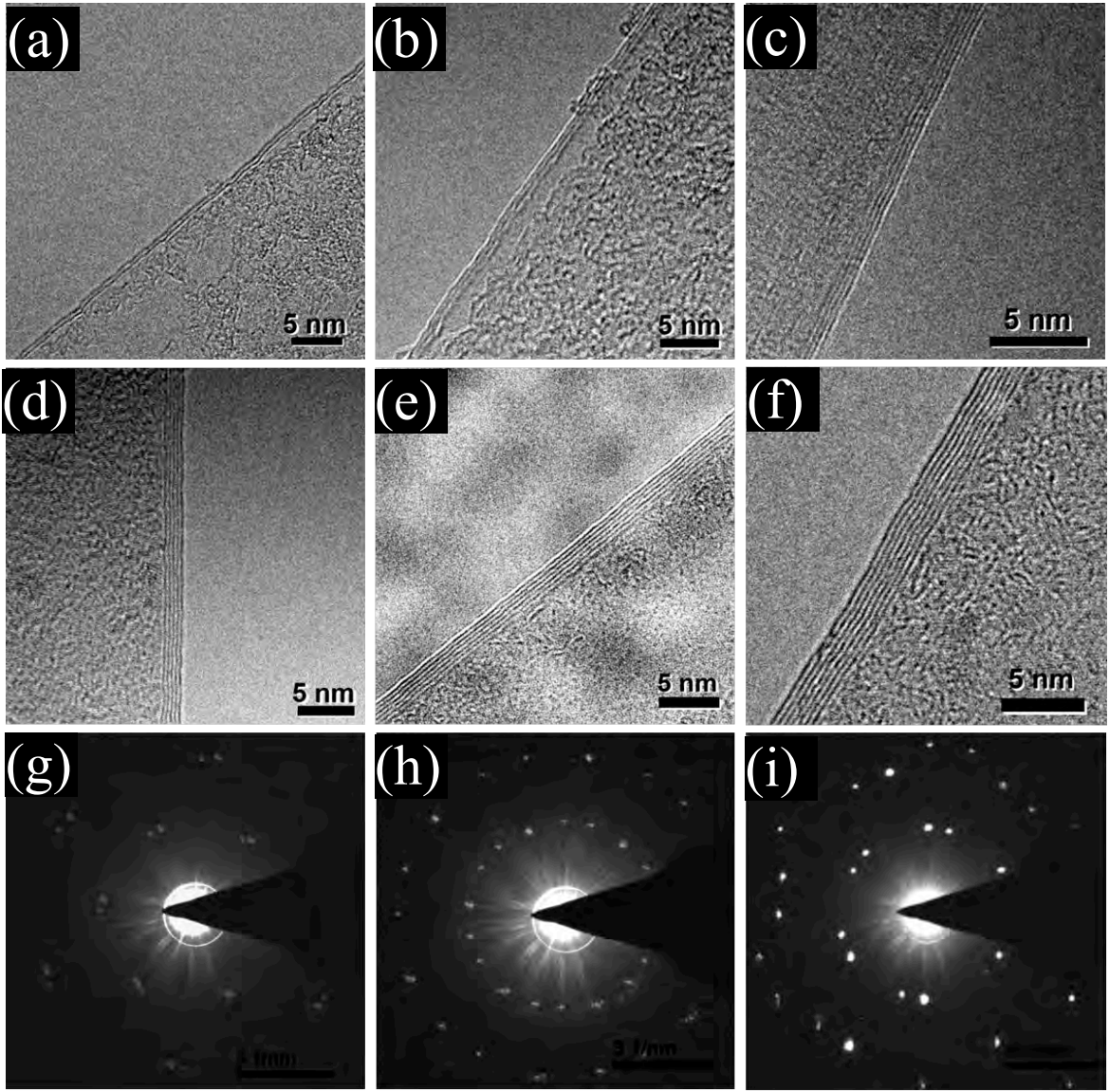

**Figure 4**

(a)
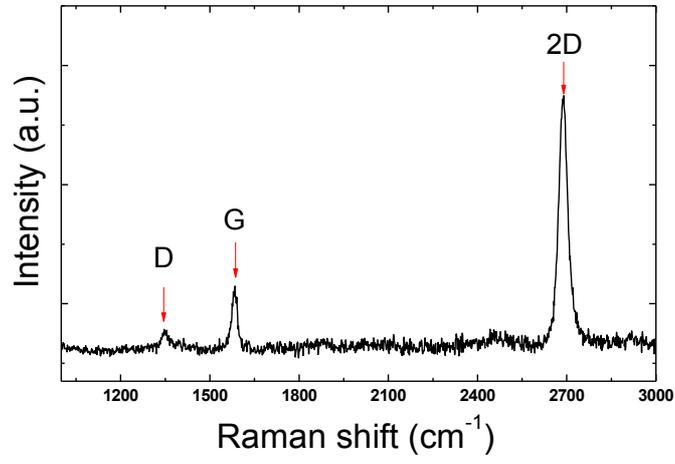

(b)
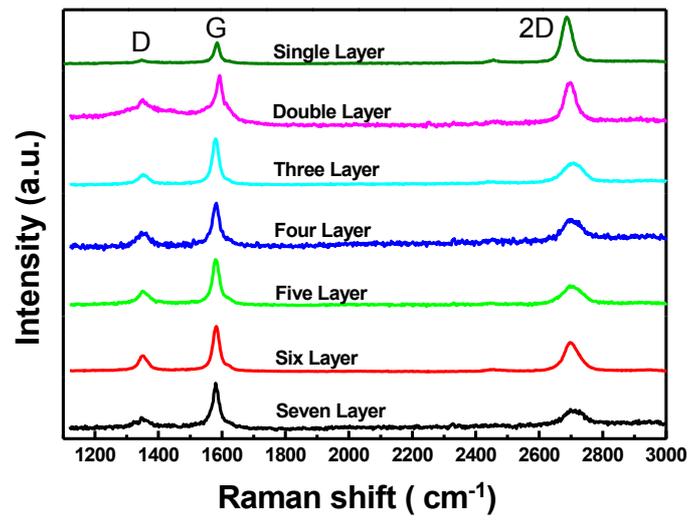

(c)
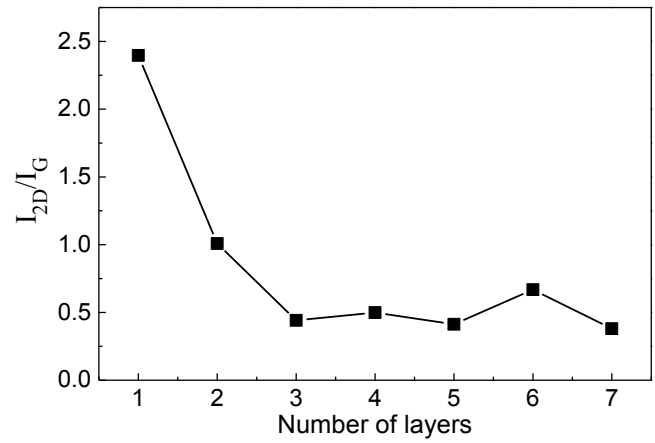

**Figure 5**

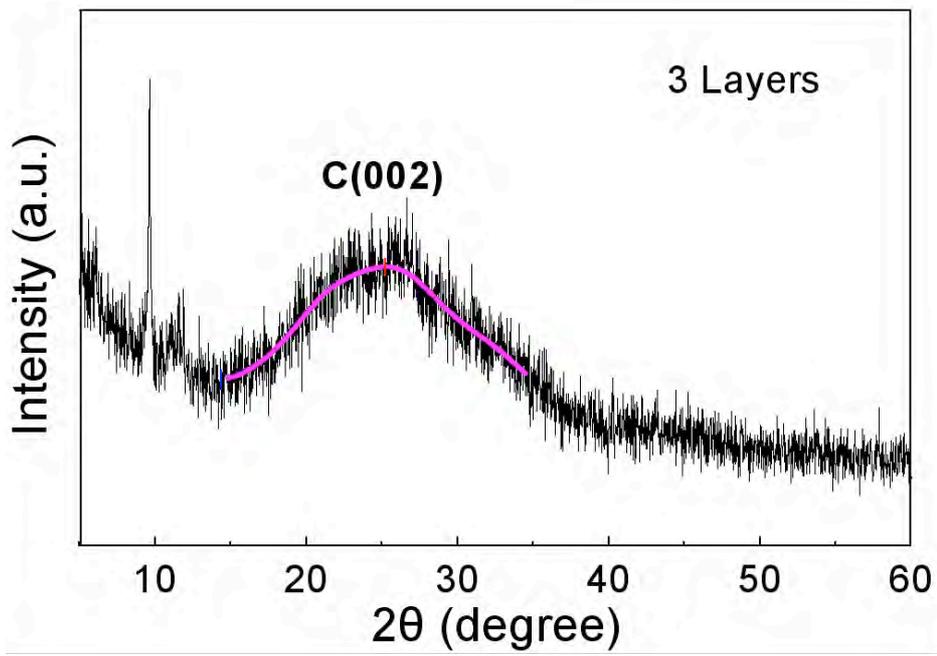

**Figure 6**

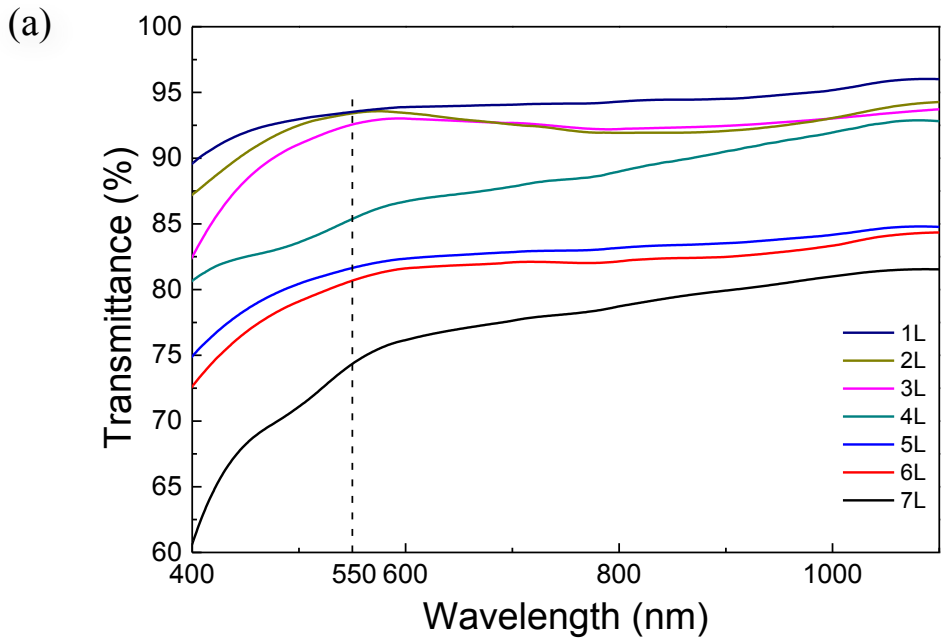

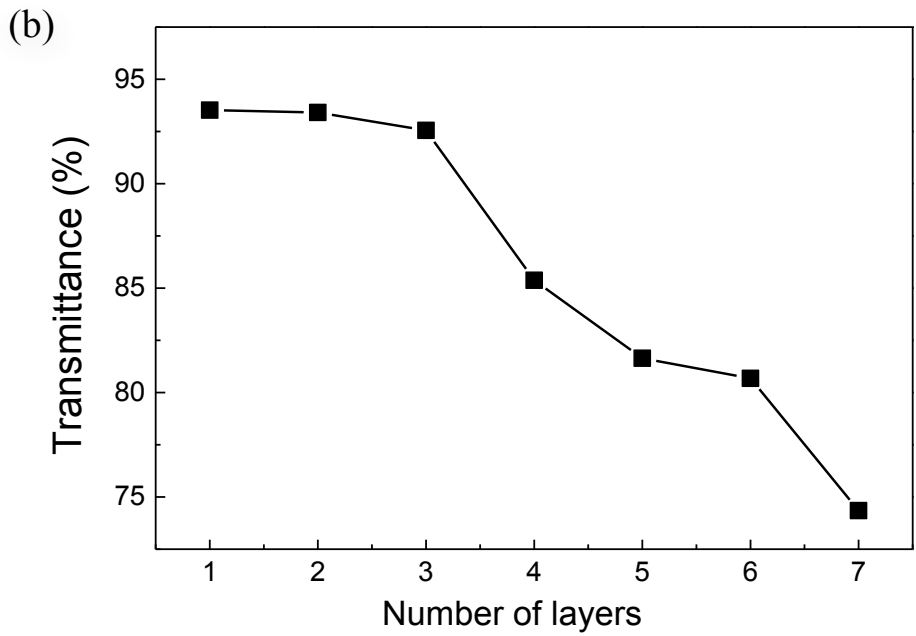